\documentclass[letterpaper, 10pt, conference]{ieeeconf}
\IEEEoverridecommandlockouts  
\usepackage{amssymb} 
\usepackage{amsmath}

\usepackage{graphicx}
\graphicspath{ {figures/} }
\usepackage{subcaption}
\usepackage{adjustbox}

\usepackage{textgreek}

\usepackage[colorlinks=true, allcolors=blue]{hyperref}

\usepackage{soul}

\usepackage{threeparttable}
\usepackage{tabularx}
\usepackage{booktabs}
\usepackage{multirow}
\usepackage{array}
\newcolumntype{Y}{>{\raggedright\arraybackslash}X}

\usepackage{xcolor}
\definecolor{lightred}{RGB}{255, 128, 128}
\definecolor{lightorange}{RGB}{255, 192, 128}
\definecolor{lightyellow}{RGB}{255, 255, 128}
\definecolor{lightgreen}{RGB}{128, 255, 128}
\definecolor{lightblue}{RGB}{128, 192, 255}
\definecolor{lightpurple}{RGB}{230, 230, 250}
\definecolor{lightpink}{RGB}{255, 192, 203}

\title{\LARGE \bf
A Graph-Based Model for Vehicle-Centric Data Sharing Ecosystem}
\author{Haiyue Yuan\textsuperscript{*}, Ali Raza\textsuperscript{\#}, Nikolay Matyunin\textsuperscript{\#}, Jibesh Patra\textsuperscript{\#}, Shujun Li\textsuperscript{*}\\
\textsuperscript{*} School of Computing, University of Kent, UK \textsuperscript{\#} Honda Research Institute Europe GmbH, Germany\\
\textsuperscript{*}\{h.yuan-221, s.j.li\}@kent.ac.uk \textsuperscript{\#}\{ali.raza, nikolay.matyunin, jibesh.patra\}@honda-ri.de%
\thanks{This is the authors' version of the accepted paper. Please cite this paper as follows: Haiyue Yuan, Ali Raza, Nikolay Matyunin, Jibesh Patra and Shujun Li (2024) A Graph-Based Model for Vehicle-Centric Data Sharing Ecosystem \emph{Proceedings of the 2024 IEEE 27th International Conference on Intelligent Transportation Systems} (ITSC 2024), pp.~3587--3594, IEEE, doi: \href{https://doi.org/10.1109/ITSC58415.2024.10919888}{10.1109/ITSC58415.2024.10919888}. For the published version, please visit the publisher's website via the DOI link.}%
}

\begin{document}
\maketitle

\begin{abstract} 
The development of technologies has prompted a paradigm shift in the automotive industry, with an increasing focus on connected services and autonomous driving capabilities. This transformation allows vehicles to collect and share vast amounts of vehicle-specific and personal data. While these technological advancements offer enhanced user experiences, they also raise privacy concerns. To understand the ecosystem of data collection and sharing in modern vehicles, we adopted the ontology 101 methodology to incorporate information extracted from different sources, including analysis of privacy policies using GPT-4, a small-scale systematic literature review, and an existing ontology, to develop a high-level conceptual graph-based model, aiming to get insights into how modern vehicles handle data exchange among different parties. This serves as a foundational model with the flexibility and scalability to further expand for modelling and analysing data sharing practices across diverse contexts. Two realistic examples were developed to demonstrate the usefulness and effectiveness of discovering insights into privacy regarding vehicle-related data sharing. We also recommend several future research directions, such as exploring advanced ontology languages for reasoning tasks, supporting topological analysis for discovering data privacy risks/concerns, and developing useful tools for comparative analysis, to strengthen the understanding of the vehicle-centric data sharing ecosystem.
\end{abstract}

\section{Introduction}
\label{sec:introduction}

With the fast development of technologies such as artificial intelligence (AI), internet of things (IoT), and 5/6G telecommunications, the automotive industry has witnessed a significant transformation towards connected vehicle services and varying degrees of autonomous driving capabilities. This has led to the integration of an increasing number of electronic control units (ECUs) and sensors within modern vehicles, empowering them to collect, process, and share a vast amount of vehicle-specific and personal data. While these capabilities offer numerous benefits such as enhanced user experience, improved safety and better efficiency, concerns regarding privacy and data security~\cite{Li-Y2023} have been raised. Referring to the well-known ``privacy paradox''~\cite{Norberg-P2007}, individuals often prioritise the functionality and convenience offered by technologies over privacy concerns, leading them to share more personal information. This would add more complexity to the emerging privacy and data security challenges in the context of vehicle-related data sharing and collection.

The extensive scale of data collection and sharing for a modern vehicle ecosystem can pose data privacy and security risks. There have been numerous related incidents reported in the real world, such as garage workers stealing and selling personal data via their' access to vehicle controller area network (CAN) bus~\cite{ico}, and massive electric vehicle (EV) driver data spilling via the usage of charging stations~\cite{Whittaker-Z2023}. Previous research has stated that a better understanding of data collection and sharing can help researchers and practitioners design better privacy and security countermeasures~\cite{Tony-N2022}, and recommended that vehicle purchasers must be informed about the full spectrum of a vehicle's data collection and sharing practices and how to properly use its privacy controls~\cite{Barber-F2023}. Having these in mind, there is a pressing need for systematic approaches to comprehend the landscape of data collection and sharing for a modern vehicle ecosystem in diverse contexts. 

In this paper, we present our work of developing a graph-based model for the vehicle-centric data sharing ecosystem, adopting the ontology 101 methodology~\cite{Noy-N2001}. We would like to address that ``data sharing'' is used here as an umbrella term to cover the whole data processing pipeline (from collection to sharing with third parties). Our methodology involves various approaches to extract key terms from publicly available materials to identify relevant entities from different perspectives of the ecosystem: 1) we adopted part of an existing ontology `VSSo'~\cite{Klotz-B2018a} to facilitate the development of our model to focus on vehicle specific signals; 2) leveraging the capabilities of a large language model (LLM), we managed to extract key terms based on the analysis of some selected modern vehicles’ privacy policies from the perspective of data sharing between organisations; and 3) we conducted a small-scale systematic literature review (SLR) to identify key entities that are related to data privacy and security of modern vehicles from a more technical perspective. This graph-based model provides high-level conceptual knowledge about how a modern vehicle collects and shares data with different parties, as well as serves as a base model for further investigation and expansion, facilitating fine-grained analysis across diverse transportation contexts.

The rest of this paper is organised as follows. Related work is discussed in Section~\ref{sec:related_work}. Section~\ref{sec:methodology} details the process of developing our graph-based model. Then Section~\ref{sec:ecosystem} presents details of the model's formal definitions in terms of its entity types and edge types. Section~\ref{sec:scenarios} applies the developed model in two real-world case studies. Section~\ref{sec:discussion} examines the limitations of this work while also proposing several future research directions. Lastly, Section~\ref{sec:conclusions} concludes this paper.

\section{Related work}
\label{sec:related_work}

Different approaches such as taxonomies, ontologies and other data models have been employed by researchers and the automotive industry to comprehend the structure and interconnections inherent in vehicle-related data. Vehicle Signal Specification (VSS)~\cite{VSS} 
introduces a domain taxonomy for vehicle signals including syntax for defining vehicle signals in a structured manner and a catalogue of vehicle-related signals. 
VSS is also the building block of the Vehicle Information Service Specification (VISS) that is under development within the W3C Automotive Working Group (\url{https://www.w3.org/TR/viss2-core/}). Based on the work of VSS, researchers and practitioners~\cite{Klotz-B2018a} developed the Vehicle Signal Specification Ontology (VSSo), later becoming part of the W3C Automotive Working Group's work. VSSo also relies on the Sensor, Observation, Sample, and Actuator (SOSA) ontology (\url{ https://www.w3.org/TR/vocab-ssn/}) for observations and actuations. Its development has evolved over the years to accommodate the latest advancement of modern vehicles, and its prime use cases include querying static/dynamic data streams for analytics purposes and supporting user interaction with vehicular data~\cite{Wilms-D2021}. Extending this work, Alvarez-Coello et al.~\cite{Alvarez-Coello-D2021} developed an ontology-based integration of vehicle-related data to understand the semantic meaning of vehicle data, such as understanding semantic descriptions of the behaviour of vehicle data streams over time and classifying dangerous driving behaviour/track locations.

Different from the above approaches that focus on understanding vehicle-specific data, Feld et al.~\cite{Feld-M2011} developed the automotive ontology, aimed at obtaining an understanding of knowledge about the users, the vehicles, and the driving situations to design next-level intelligent in-car systems for better-managing knowledge inside a vehicle and sharing these knowledge between vehicles. The core of the ontology comprises a user model and a context model, where the former focuses on users' preferences and interactions, and the latter addresses aspects related to the vehicle, trips and in-car devices. Moreover, ontological models have been adopted to support driving decision-making and autonomous driving. Zhao et al.~\cite{Zhao-L2017} utilised map, control, and car ontological models for translating the sensor data in a machine-understandable format to develop a driving decision-making system that allows a vehicle to understand maps and driving paths/environments to make safety decisions in real-world driving. Fernandez et al.~\cite{Fernandez-S2016} introduced an ontology to represent different concepts involved in the road traffic scenario and developed a system that combines the information provided by a traffic sensor network with ontology-based knowledge bases to improve the driving environment. Slightly differently, Viktorovic et al.~\cite{Viktorovic-M2020} proposed the Connected Traffic Data Ontology (CTDO) based on the SOSA ontology to support a network of connected vehicles.

Despite the plethora of research mentioned above, previous studies have not systematically looked into the following: 1) what main parties are involved in data sharing for a modern vehicle; 2) how data flows between different parties; and 3) what insights about data privacy would such data flows reveal. By introducing the proposed graph-based model, we hope to fill these research gaps.

\section{Methodology}
\label{sec:methodology}

We adopted the ``ontology development 101'' method~\cite{Noy-N2001}, one of the most well-known methodologies for guiding individuals in ``how to model'' a domain by selecting constructs and entities, to derive our graph-based model. The rest of this section details how we adapted the 7 phases of the method to develop our graph-based model.

\subsubsection{Phase 1 -- Determine the domain and scope}
\label{sec:phase1}

The main objective here is to determine the domain and scope of our work. We chose the automotive industry as the domain and the vehicle-centric data sharing ecosystem as the scope.

\subsubsection{Phase 2 -- Consider reusing existing ontologies/models}
\label{sec:phase2}

Among existing ontologies, we decided to adopt VSSo~\cite{Wilms-D2021}. It is based on VSS, which includes over 1,500 distinct vehicle components. Explicitly representing each of them in our model would greatly amplify its complexity and the challenges associated with visualisation. Hence, we did not dig deep into the fine details of VSSo, instead, we adopted its primary components as the entities within our model, specifically focusing on \emph{Vehicle} and \emph{Vehicle Component} entities. The semantic relation between these two is \emph{isPartOf}. This relation is pertinent between a vehicle and its components and applies recursively among \emph{Vehicle Components} themselves to accurately represent their hierarchical structure. We also used a general data sharing graphical model proposed by Lu \& Li~\cite{LL2022PrivacyBenefitModel} to inform our proposed model.

\subsubsection{Phase 3 -- Enumerate important terms}
\label{sec:phase3}

\begin{table}[!htb]
\centering
\begin{threeparttable}
\caption{Data sharing destinations extracted with GPT-4 from privacy policies, with manually assigned labels}
\label{tab:sharing_destinations}
\begin{tabularx}{\linewidth}{c X}
\toprule
Brand & Extracted personal data sharing destinations\\
\midrule
Ford & IT service provider\textsuperscript{\textalpha}, HERE Global B.V\textsuperscript{\textalpha}, Vodafone\textsuperscript{\textalpha}, Garmin\textsuperscript{\textalpha}, Digital Roadside Assistance\textsuperscript{\textalpha}, Ford Secure\textsuperscript{\textbeta}, Ford Smart Mobility UK Limited\textsuperscript{\textbeta}, Authorised dealer\textsuperscript{\texttheta}, Law enforcement\textsuperscript{\textgamma}, New business owner\textsuperscript{\textzeta}\\
\midrule
Tesla & Service providers\textsuperscript{\textalpha}, Business partners\textsuperscript{\textdelta}, Payment processors\textsuperscript{\textalpha}, Financial institutions\textsuperscript{\textalpha}, Energy utilities service provider\textsuperscript{\textalpha}, Affiliates and subsidiaries\textsuperscript{\textbeta}, Law enforcement\textsuperscript{\textgamma}, Government authorities\textsuperscript{\textgamma}, Marketing partners\textsuperscript{\textdelta}, Third-party service and repair centres\textsuperscript{\texteta}\\
\midrule
Renault & Third-party service providers\textsuperscript{\textalpha}, Third parties for legal obligations\textsuperscript{\textalpha}, Approved dealer\textsuperscript{\texttheta}, Other companies in Groupe Renault\textsuperscript{\textbeta}, Renault SAS\textsuperscript{\textbeta}, Business partners\textsuperscript{\textalpha}, Law enforcement\textsuperscript{\textgamma}, Courts\textsuperscript{\textgamma}, Government and tax authorities\textsuperscript{\textgamma}, Social media companies\textsuperscript{\textalpha}\\
\midrule
Nissan & Various service providers\textsuperscript{\textalpha}, Service partners\textsuperscript{\textdelta}, Third-party service providers\textsuperscript{\textalpha}, Nissan-Affiliated companies\textsuperscript{\textbeta}, Public authorities and courts\textsuperscript{\textgamma}, Third Parties in business transfers\textsuperscript{\textzeta}\\
\midrule
Honda & e3 Media (Great State)\textsuperscript{\textalpha}, Third-party hosting providers\textsuperscript{\textalpha}, Professional advisors\textsuperscript{\textalpha}, Sub-Contractors\textsuperscript{\textalpha}, Worldline\textsuperscript{\textalpha}, SoundHound Inc.\textsuperscript{\textalpha}, Concentrix\textsuperscript{\textalpha}, Snap-On\textsuperscript{\textalpha}, Bosch Service Solutions GmbH\textsuperscript{\textalpha}, IBM\textsuperscript{\textalpha}, ICUC\textsuperscript{\textalpha}, Digitalist Group Plc.\textsuperscript{\textalpha}, Companies within Honda Group\textsuperscript{\textbeta}, Regulators\textsuperscript{\textgamma}, Law enforcement\textsuperscript{\textgamma}, Courts\textsuperscript{\textgamma}, HMRC and other tax bodies\textsuperscript{\textgamma}, Business partners\textsuperscript{\textdelta}, Other marketing partners\textsuperscript{\textdelta}, Prospective business buyers\textsuperscript{\textzeta}, Asset acquirers\textsuperscript{\textzeta}\\
\bottomrule
\end{tabularx}
\begin{tablenotes}
\item \textalpha: \emph{Third party company}, \textbeta: \emph{Affiliated company}, \textdelta: \emph{Business partner}, \texttheta: \emph{Dealer}, \textgamma: \emph{Government body}, \textzeta: \emph{Business buyer}, \texteta: \emph{Service centre}
\end{tablenotes}
\end{threeparttable}
\end{table}

\begin{table}[!htb]
\centering
\caption{Search query we used for identifying relevant research papers using Scopus}
\label{tab:search_query}
\begin{tabularx}{\linewidth}{c X}
\toprule
AND-term & Keyword combination(s)\\
\midrule
Data privacy & \texttt{"data privacy" OR "privacy" OR "data security" OR "security"}\\
Vehicle & \texttt{"connected vehicle" OR "electric vehicle" OR "vehicle" OR "autonomous vehicle"}\\
Survey & \texttt{"survey" OR "review" OR "systematic review" OR "systematic study"}\\
\bottomrule
\end{tabularx}
\end{table}

We used two additional methods to facilitate the identification of relevant terms. \textbf{Method 1: the use of an LLM.} We adopted a few-shot learning approach to develop a customised model of the LLM GPT-4 (\url{https://openai.com/research/gpt-4}), which is capable of processing large volumes of text and producing structured output in JSON format. This output resembles entity-relation data, consisting of the types of data shared, the intended data sharing purposes, and the recipients (i.e., entities) of the shared data. We took the privacy policies of several selected car brands and fed them into the LLM for automatic analysis. We managed to derive some key terms of data sharing destinations as summarised in Table~\ref{tab:sharing_destinations}\footnote{The LLM is unlikely to be able to comprehend the complicated data sharing landscape, but it can help derive candidate key entities for our model.}. Then we carefully reviewed all terms and manually grouped those with similar meanings and assigned distinct labels denoted by various superscript symbols as illustrated in Table~\ref{tab:sharing_destinations}. \textbf{Method 2: the use of SLR.} We conducted a small-scale SLR using Scopus (\url{https://www.scopus.com/}), following the search query shown in Table~\ref{tab:search_query}. The search was done in February 2024, and applied to meta-data (title, abstract and keywords). We focused on survey/review papers within the disciplines of computer science and engineering, written in English, and published in the past 12 months. Out of the 91 returned results, 42 were excluded based on the following reasons: they focused on other research areas such as unmanned aerial vehicles and maritime vehicles; and they were not review papers. By examining all remaining 49 articles, we further narrowed down to 13 articles that cover vehicle-related data privacy and security from the perspective of ecosystem or network, e.g., vehicle-to-everything (V2X), vehicle-to-network (V2N), and vehicle-to-grid (V2G). Lastly, we extracted key terms that involve data sharing and data privacy of vehicles, and the results are depicted in Table~\ref{tab:key_terms_LR}.

\begin{table}[!htb]
\begin{threeparttable}
\centering
\caption{Key terms identified using the small-scale SLR}
\label{tab:key_terms_LR}
\begin{tabularx}{\linewidth}{X c}
\hline
Terms & Papers\\
\hline
\sethlcolor{lightpink}\hl{Road Side Unit (RSU)}, \sethlcolor{lightpurple}\hl{On Board Unit (OBU)}, \sethlcolor{lightyellow}\hl{Network Infrastructure, Satellite} & \cite{Yoshizawa-T2023}\\
\hline
\sethlcolor{lightpink}\hl{RSU} ,\sethlcolor{lightpurple}\hl{OBU}, Vulnerable Road Users (VRU), \sethlcolor{lightyellow}\hl{Base Station} & \cite{Sedar-R2023}\\
\hline
\sethlcolor{lightred}\hl{Charging Station, Charging Spot, Smart Meter} & \cite{Arun-S2023}\\
\hline
\sethlcolor{lightpurple}\hl{OBU}, \sethlcolor{lightpink}\hl{RSU} & \cite{Al-Shareeda-M2023}\\
\hline
\sethlcolor{lightpurple}\hl{Electronic Control Unit (ECU), CAN}, \sethlcolor{lightpink}\hl{RSU}, \sethlcolor{lightred}\hl{Power Station}, \sethlcolor{lightgreen}\hl{Devices (i.e., mobile phones, tablet)}, \sethlcolor{lightyellow}\hl{Base Station} & \cite{Samira-T2023}\\
\hline
\sethlcolor{lightpurple}\hl{ECU, CAN}, \sethlcolor{lightpink}\hl{RSU} & \cite{Lampe-B2023}\\
\hline
\sethlcolor{lightpurple}\hl{ECU, CAN, Sensors}, \sethlcolor{lightblue}\hl{LiDAR, RADAR}, \sethlcolor{lightyellow}\hl{Satellite} & \cite{Qurashi-J2023}\\
\hline
\sethlcolor{lightpink}\hl{RSU Infrastructure}, \sethlcolor{lightblue}\hl{Sensor}, \sethlcolor{lightyellow}\hl{Cloud Server}, \sethlcolor{lightgreen}\hl{Personal Device} & \cite{Nazia-T2023}\\
\hline
\sethlcolor{lightpurple}\hl{ECU, CAN, Wifi, Bluetooth}, \sethlcolor{lightpink}\hl{RSU}, \sethlcolor{lightblue}\hl{Sensors}, \sethlcolor{lightyellow}\hl{Cellular, Satellite} & \cite{Qurashi-J2023a}\\
\hline
\sethlcolor{lightyellow}\hl{GPS}, \sethlcolor{lightblue}\hl{LiDAR, RADAR, Cameras}, \sethlcolor{lightpink}\hl{Traffic Lights} & \cite{Rana-K2023}\\
\hline
\sethlcolor{lightpink}\hl{RSU}, \sethlcolor{lightpurple}\hl{OBU, Advanced Driver Assistance Systems (ADAS)}, \sethlcolor{lightyellow}\hl{Base Station} & \cite{Mehta-A2024}\\
\hline
\sethlcolor{lightred}\hl{Charging Infrastructure, Charging Port} & \cite{Ronanki-D2024}\\
\hline
\sethlcolor{lightpink}\hl{RSU}, \sethlcolor{lightpurple}\hl{OBU}, \sethlcolor{lightorange}\hl{Enforcement System (Cameras)} & \cite{Adavoudi-J2023}\\
\hline
\multicolumn{2}{c}{Colour codes: \sethlcolor{lightpurple}\hl{Defined in VSSo}, \sethlcolor{lightyellow}\hl{Network infrastructure},\sethlcolor{lightgreen}\hl{Digital asset},\sethlcolor{lightpink}\hl{RSU},} \\
\multicolumn{2}{c}{\sethlcolor{lightred}\hl{Charging facilitiy},\sethlcolor{lightblue}\hl{Additional vehicle sensor},\sethlcolor{lightorange}\hl{Traffic monitoring sensor}} \\
\end{tabularx}
\end{threeparttable}
\end{table}

\subsubsection{Phase 4 -- Define the classes (entities) and the hierarchy}

In addition to the two entity types (i.e., `Vehicle' and `Vehicle component') defined in Section~\ref{sec:phase2}, we took additional steps to further finalise key entity types from the extracted terms listed in Section~\ref{sec:phase3}. Referring to Table~\ref{tab:sharing_destinations}, many of the terms such as `\emph{Third party company}', `\emph{Business partner}', `\emph{Affiliated company}' extracted from the privacy policies are considered as parties that provide various services to the vehicle and the vehicle's users. Taking into account their similarities, we consolidated them under the overarching key entity type `Service provider'. However, due to the administrative/regulatory nature of the `Government body', we would keep it as a separate entity type in the ecosystem. Regarding terms identified in the small-scale SLR, we grouped terms with similar meanings. As shown in Table~\ref{tab:key_terms_LR}, terms highlighted in light purple (e.g., OBU, ECU, CAN, Wifi, Bluetooth, and ADAS) are part of VSSo and considered as entity type `Vehicle components'. Then, the final list of key entity types derived from the small-scale SLR includes `Network infrastructure', `RSU', `Digital asset', `Charging facility', `Additional vehicle sensor', and `Traffic monitoring sensor'. Moreover, we introduce four self-developed entity types, namely `Person', `Organisation', `Data Package', and `Communication infrastructure' to enrich the model. ``Organisation'' represents entities that may have connections to other entities for various purposes. Due to the similar characteristics, here we consider `Government body' and `Service provider' as subclasses of `Organisation'. `Data package' refers to a specific combination of atomic data items that would pass between properties. Additionally, considering the similar properties of `RSU' and `Network infrastructure', we categorised them as sub-classes of the entity type `Communication infrastructure'.

\subsubsection{Phases 5 \& 6 -- Define the properties of classes-slots \& Define the facets of the slots}

Here we merge the two phases of the Ontology Development 101 methodology as they are intertwined. From our SLR and other papers we found ad-hoc searches~\cite{Mu-H2021, Lee-C2023}, privacy preservation has been frequently regarded as a potential solution to address privacy issues for various entity types. Here we specifically focus on privacy implications in the data sharing ecosystem, therefore we identified entity types that have been subject to privacy preservation discussion and subsequently added `\emph{privacy preserving}' as an attribute to these entity types. Moreover, we adopted two attributes introduced in VSSo to entity types `Vehicle' and `Vehicle component'. One is `\emph{static property}' that corresponds to the \emph{StaticVehicleProperty} in VSSo\footnote{It refers to a particular characteristic of a vehicle or vehicle component such as the vehicle's height, length and VIN (vehicle identification number)}. Another attribute is `\emph{dynamic property}', which corresponds to the \emph{DynamicVehicleProperty} defined in VSSo\footnote{It represents a signal that is continuously changing over time such as the vehicle's speed and acceleration}. These attributes in VSSo are specifically designed to represent various vehicle-specific signals/data, we also intended to capture them in our proposed model, as some of them may be considered as sensitive information, particularly when combined with other data sources to infer more detailed information. Furthermore, we would like to introduce uni-directional and bi-directional edges to model the direction in which data may flow between different entity types.

\subsubsection{Phase 7 -- Create instances}

This phase involves creating individual instances of the graph-based model to assess its applicability. We produced two use cases demonstrating the model's usefulness and effectiveness, and further details can be found in Section~\ref{sec:scenarios}.

\begin{figure*}[!htb]
\centering
\includegraphics[width=.7\linewidth]{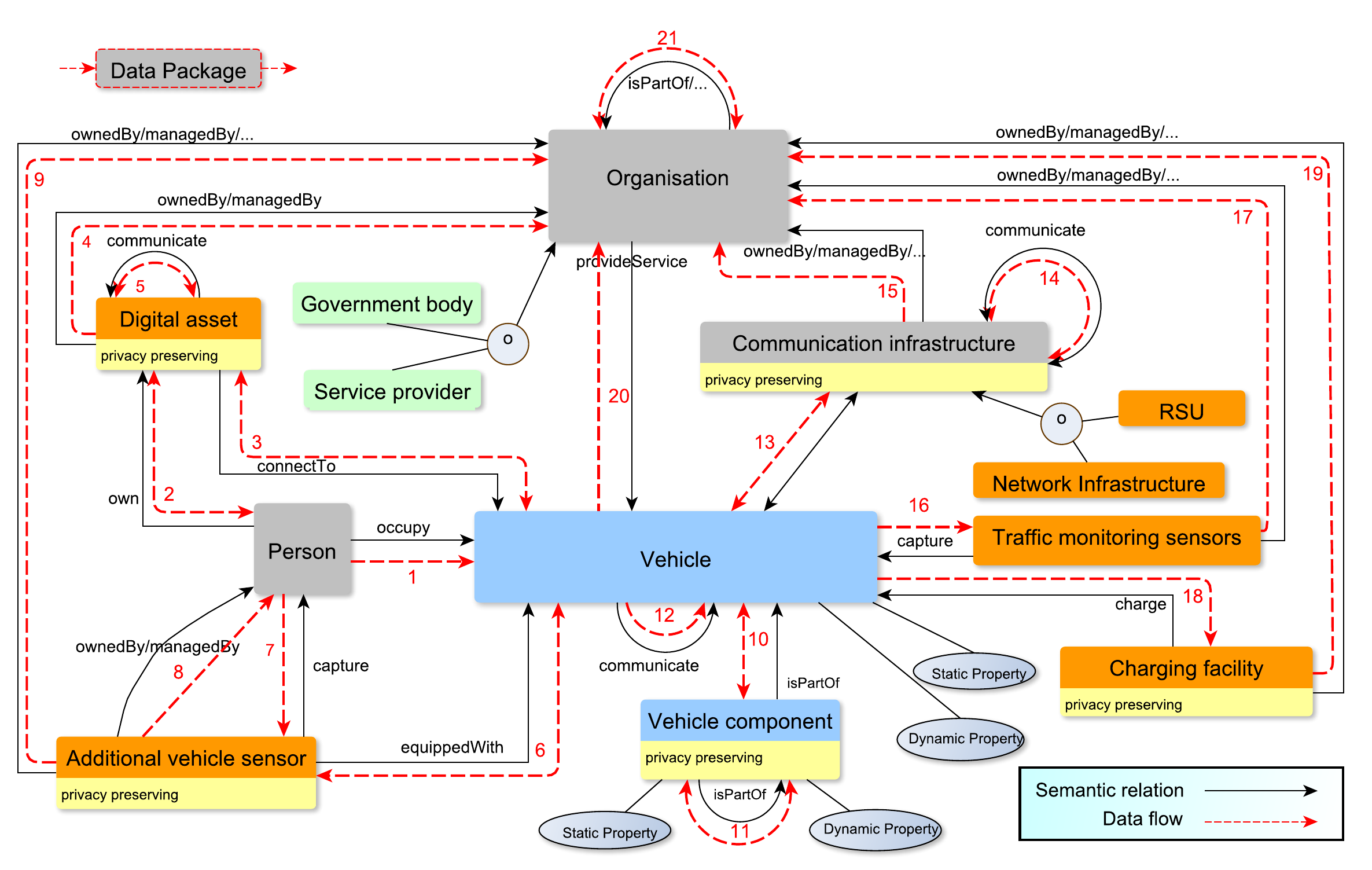}
\caption{An entity type graph}
\label{fig:graph_model}
\end{figure*}

\section{The graph-based model} 
\label{sec:ecosystem}

The graph-based model can be formalised as a directed graph as shown in Figure~\ref{fig:graph_model}, describing how data can potentially flow through different types of entities. The graph can be formally described as $\mathcal{G} = (\mathcal{V}, \mathcal{E})$, where $\mathcal{V}={\{\mathcal{V}_i}\}^M_{i=1}$ and $\mathcal{E}={\{\mathcal{E}_j}\}^N_{j=1}$ represent a set of $M$ nodes and a set of $N$ edges, respectively. Each node $\mathcal{V}_i$ represents one entity type that is depicted by a rounded corner rectangle in the proposed graph model. Edges in $\mathcal{G}$ can be classified into two types: semantic relations (i.e., solid line) and data flow (i.e., red dashed line). Such an entity type graph $\mathcal{G}$ provides a high-level representation of entity types and relations among them. Our method of using different sources for entity identification enabled us to cover different levels of abstraction. The analysis of privacy policies covers the high-level data flows between organisations, while the small-scale SLR and the adoption of VSSo complete the work with a focus on technical aspects of the ecosystem. However, the entity type graph does not capture the specific entities and relationships. 
It is necessary and important to investigate the vehicle-centric data sharing ecosystem in greater depth at the entity level. For this purpose, we further defined entity-level graphs, where each of such graphs is represented as a directed graph $G = (\mathbb{V},\mathbb{E})$. It consists of a set of instance nodes $\mathbb{V}=\{v|v\in\mathcal{V}_i, 1 \leq i \leq M\}$, where each node represents an instance entity (i.e., an instance of a specific entity type/node in $\mathcal{G}$), and a set of instance edges $\mathbb{E}=\{e|e \in \mathcal{E}_i, 1 \leq j \leq N\}$, where each instance edge represents an instance of a specific relation type/edge in $\mathcal{G}$.

\subsection{Entity types}
\label{sec:entity}

Entity types presented in this section are colour-coded, where blue represents entities derived from existing ontologies/data model, orange entities are extracted from the SLR, entities in green are the outcome of the privacy policy analysis using the GPT model, and entities in grey are self-developed. We explicitly retain specific subclass entity types in the model for two reasons: 1) they are directly derived from either SLR or privacy policy analysis; and 2) they can enrich the model with additional context. In the following, we present more details of all entity types in alphabetic order.

\textbf{Additional vehicle sensor (AVS)}: a sensor or a sensing system installed in a vehicle to gather data related to the vehicle's operation, environment, or occupants.

\textbf{Vehicle (V)}: a means of transport (vehicle) designed to carry passengers and/or goods.

\textbf{Vehicle component (VC)}: an individual part/element of a vehicle.

\textbf{Charging facility (CF)}: an infrastructure designed to provide battery charging services for electric vehicles.

\textbf{Communication infrastructure (CI)}: infrastructure that enables vehicle-related communication between entities in the ecosystem. Two subclass entity types are: 1) \textbf{Network infrastructure (NI)} refers to infrastructures or equipment designed to facilitate network communication and connectivity; and 2) \textbf{Road side unit (RSU)} refers to gateways between vehicles' OBUs and the communication infrastructure.  

\textbf{Digital asset (DA)}: an electronic device (i.e., mobile phone) or a digital service (i.e., mobile app) that can be connected to a vehicle for communication, entertainment, and assisting driving.

\textbf{Data package (DP)}: a collection of data items that are transmitted/shared between two entity types for one or more specific purposes. Here one data item refers to one piece of data in its atomic format.

\textbf{Organisation (O)}: an organisation that relates to one or more other entities in the ecosystem. Two subclass entity types are: 1) \textbf{Government body (G)} refers to an  organisation or entity established by a government or governing authority to carry out specific functions and or duties; and 2) \textbf{Service provider (SP)} refers to an organisation that offers a specific service to vehicles.

\textbf{Person (P)}: an individual human being, who can use the vehicle as either a driver or a passenger.

\textbf{Traffic monitoring sensor (TMS)}: a device or a system designed to monitor and manage traffic conditions.

\subsection{Edge types}
\label{sec:edge}

We develop two main edge types for this model: 1) `Semantic relation' is denoted by a solid directed line with accompanying text labels, aiming to model how and why data may flow between two entities; and 2) `data flow' is depicted by a dashed red line, and the associated arrow indicates the direction of the data flow. As shown in Figure~\ref{fig:graph_model}, we choose not to explicitly include DP entities within the graph for better visual representation. Instead, a disconnected single DP with one dashed red line pointing towards it and another pointing away is used to indicate the co-existence of a DP entity with any data flows. We will use $E_i$ to denote a unique data flow edge type between two entity types.

As shown in Figure~\ref{fig:graph_model}, a semantic relation `\emph{occupy}' is used to describe the relation between P and V entities. We consider an `\emph{occupy}' relation to have two different semantic meanings: 1) a person \emph{occupies} a vehicle as a driver; and 2) a person \emph{occupies} a vehicle as a passenger. The data flow edge denoted by $E_1$ with a uni-directional arrow indicates that data (e.g., driving behaviour and voice data) can flow from P to V. In theory, entity types such as NI, CF and RSU should directly interact with specific vehicle components. However, for the sake of simplicity, we choose not to depict such relations in the graph to avoid overly complicating the representation on a low/technical level. Instead, we consider VC entities as enablers of such relations through the semantic relation `\emph{isPartOf}' between V and VC entities. A person can connect their digital assets (e.g., mobile phones and mobile apps) to a vehicle for add-on services such as assisted driving and entertainment. Personal data could be collected by the vehicle via its connection with the connected digital assets. This is modelled as a bi-directional edge type $E_2$ and $E_3$ in our model. Considering that a digital asset such as a mobile app can be `\emph{ownedBy/managedBy}' an organisation, its data could be accessible by the organisation. We denote edge type $E_4$ to model such data flows. Moreover, edge type $E_5$ is used to describe data flows when one digital asset \emph{communicate} with another digital asset while multiple persons are involved.

Furthermore, additional sensors may be installed on vehicles as extra road safety measures (e.g., a cabin-facing dashcam on a taxi and CCTV cameras on a bus) or to facilitate autonomous driving (e.g., LiDAR/RADAR). The relation `\emph{equippedWith}' describes such semantic relations between entity types AVS and V and $E_6$ is used to represent this bi-directional data flow edge. The edge type $E_7$ represents the flow of personal data to an AVS. Depending on the ownership of the installed sensors, edge types $E_8$ and $E_9$ represent the data flows between AVS and P entities and between AVS and O entities, respectively. The relation `\emph{isPartOf}' represents the semantic relation between V and VC entities, and between two VC entities. All modern vehicles, whether traditional combustion-powered cars or the latest EVs, are equipped with a CAN bus, which connects a large number of ECUs to facilitate data transmission among various components of the vehicle. Such data flows are modelled and denoted by edge type $E_{10}$ and $E_{11}$ in our model.

Data exchanges between network infrastructures and RSUs enable real-time information exchanges, contributing to safer and more efficient travel. In this model, we use $E_{12}$ to represent data communication among multiple vehicles. $E_{13}$ and $E_{14}$ represent the bi-directional data flows between V and CI entities and between different CIs entities (e.g., data flows between RSU and NI), respectively. Edge type $E_{15}$ describes the data flow from CI to O entities. Alongside CIs, traffic monitoring sensors (e.g., speed cameras, and automatic number plate recognition (ANPR) cameras) also play a crucial role in ensuring road safety. Differently, traffic monitoring sensors capture data in a passive approach, and edge type $E_{16}$ models such data flows between V and TMS entities. Similar to the entity type CI, the ownership of TMS is vital in determining the direction of data flow as denoted as edge type $E_{17}$. Moreover, previous research~\cite{Tony-N2022} has indicated that large amounts of data could be leaked from EVs while charging. This is modelled using edge type $E_{18}$ between CF and V entities. The ownership of a charging facility is important to data privacy, especially when third-party charging services are involved. We use $E_{19}$ to model data flows between CF and O entities. Furthermore, we use the semantic relation `\emph{provideService}' to represent various services that an organisation can provide to a vehicle while a wide range of data can be shared in exchange compulsorily, voluntarily, or optionally. Here, the edge type $E_{20}$ describes data flows between O and V entities. Additionally, an organisation may be affiliated with another organisation in different capacities such as a subsidiary and a business partner. The potential route of sharing data with these affiliated companies has been explicitly addressed in all analysed privacy policies. This is captured using the edge type $E_{21}$ in our model.

\section{Modelling real-world scenarios}
\label{sec:scenarios}

In this section, realistic examples are used to develop entity-level graphs. The graphs presented in this section will not display the entire entity-type graph, but only focus on exploring specific sub-graphs to highlight relevant data flows.

\subsection{Data flows for booking an Uber car with a dashcam}
\label{sec:uber}

\begin{figure}[!htb]
\centering
\includegraphics[width=\linewidth]{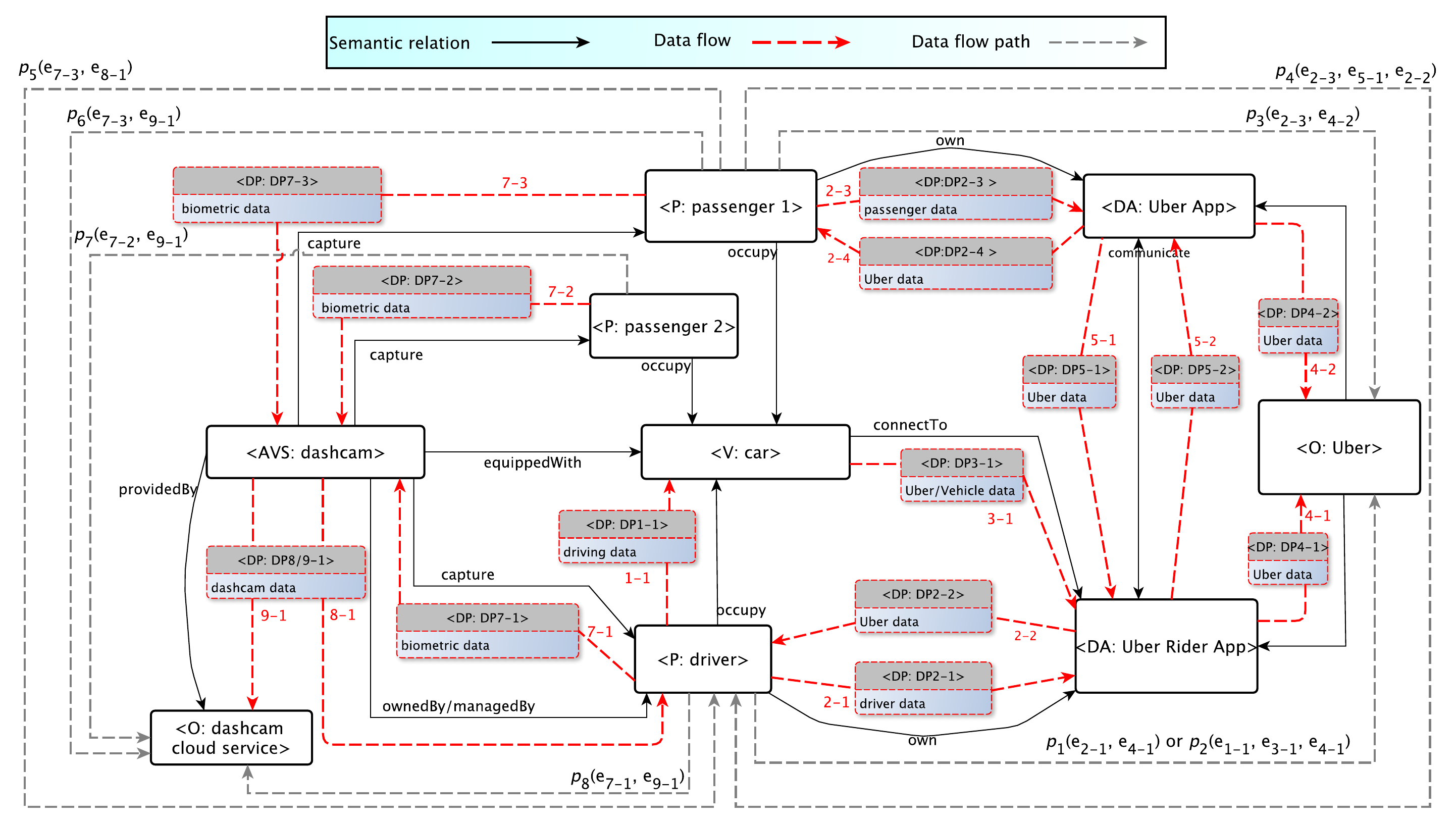}
\caption{An entity-level graph for an Uber booking scenario}
\label{fig:taxi_model}
\end{figure}

In this part, we present our work of generating an entity-level graph to model a use case of two persons travelling in a booked Uber car equipped with a dashcam. A simple entity-level graph involving P, V, DA, AVS, and O entities is presented in Figure~\ref{fig:taxi_model}. Since we don't know exactly what data packages would be transmitted between entities, we name each data package following the pattern of 'DP + ${e}_i$' with a short text description underneath the DP entity. This is also applied to the use case 
in Section~\ref{sec:speeding}.

\subsubsection{Modelling}

A $<$driver$>$ drives the $<$car$>$ as denoted as \emph{occupy} semantic relation between two entities, resulting in the flow of data package $<$DP1-1$>$, consisting of driving data such as driver's driving habits and driving speed. This is modelled using edge type $E_1$, as denoted by $e1_1$. The Uber $<$driver$>$ uses $<$Uber Rider app$>$ to manage and receive trip bookings, which causes the flow of $<$DP2-1$>$ (e.g., the number plate, the car make and the model, the driver's name) from $<$driver$>$ to $<$Uber Rider app$>$, denoted by $e_{2-1}$. The $<$driver$>$ connects the $<$Uber Rider app$>$ to the $<$car$>$ leading the data exchange (i.e., $<$DP3-1$>$ ) between both entities, denoted by $e_{3-1}$. $<$passenger 1$>$ uses an $<$Uber app$>$ to book a car, resulting in the flow of $<$DP2-3$>$ (e.g., location data, destination data, the passenger's name and phone number) from $<$passenger 1$>$ to $<$Uber app$>$, denoted by $e_{2-3}$. During the booking process, both $<$Uber app$>$ and $<$Uber Rider app$>$ communicate with each other, leading the exchange of $<$DP5-1$>$ and $<$DP5-2$>$ between them. This is modelled using edge type $E_5$, denoted by $e_{5-1}$ and $e_{5-2}$ respectively. In addition, both $<$Uber app$>$ and $<$Uber Rider app$>$ are products of $<$Uber$>$, thereby, $<$DP4-1$>$ and $<$DP4-2$>$ will be shared with the organisation for various purposes. These are modelled using $E_4$ edge type, denoted by $e_{4-1}$ and $e_{4-2}$. As shown in Figure~\ref{fig:taxi_model}, a $<$dashcam$>$ is installed and managed by the $<$driver$>$ for both driver and passengers' safety. The $<$dashcam$>$ collects biometric data (i.e., $<$DP7-1$>$, $<$DP7-2$>$, $<$DP7-3$>$) such as image and voice data from all three persons, and such data flows are modelled using $E_7$ edge type, and denoted by $e_{7-1}$, $e_{7-2}$, $e_{7-3}$, respectively. The driver has access to all recorded data (i.e., $<$DP8/9-1$>$), this is modelled using $E_8$ edge type, denoted by $e_{8-1}$. Similarly, the organisation $<$dashcam cloud service$>$ provides service to the $<$dashcam$>$, meaning that recorded data can be available to the organisation, and $e_{9-1}$ is used to model such a data flow.

\subsubsection{Path analysis and discussion}

As illustrated by grey dashed lines, there are two possible paths (i.e., $p_1=(e_{2-1}, e_{4-1})$ and $p_2=(e_{1-1}, e_{2-1}, e_{4-1}$) for $<$Uber$>$ to get the $<$driver$>$'s data. $<$Uber$>$ can also get $<$passenger 1$>$'s data via the path $p_3=(e_{2-3}, e_{4-2})$. In addition, $<$driver$>$ can get $<$passenger 1$>$'s data via $p_4=(e_{2-3}, e_{5-1}, e_{2-2})$. On the dashcam side, $<$dashcam$>$ captures biometric data such as images and voices of persons in the cabin, and shares the data with the organisation $<$dashcam cloud service$>$, which stores data from $<$passenger 1$>$, $<$passenger 2$>$, and $<$driver$>$ via path $p_6=(e_{7-3}, e_{9-1})$, $p_7=(e_{7-2}, e_{9-1})$, and $p_8=(e_{7-1}, e_{9-1})$ respectively. Since the driver is the owner of the dashcam, the driver has access to the data $<$DP8/9-1$>$ which is the collection of all persons' data, so apart from the path $p_4$, the $<$driver$>$ can also get $<$passenger 1$>$'s data via path $p_5=(e_{7-3}, e_{8-1})$. As illustrated here, our approach can reveal complex data sharing dynamics between different entity types. Both the driver and Uber would have access to one person's data from different sources, and the ability of data aggregation to infer more sensitive information could lead to escalated data privacy concerns. As data sharing becomes more prevalent, the responsibility falls on both individual and the organisation to implement effective data security and privacy protection strategies. In response to emerging demands, such entity-level graph can help researchers/practitioners to analyse real-world cases systematically and derive all possible data flow paths for facilitating developing appropriate solutions (e.g., access control and privacy preservation schemes).

\subsection{Data flows in speeding incident}
\label{sec:speeding}

\begin{figure}[!htb]
\centering
\includegraphics[width=\linewidth]{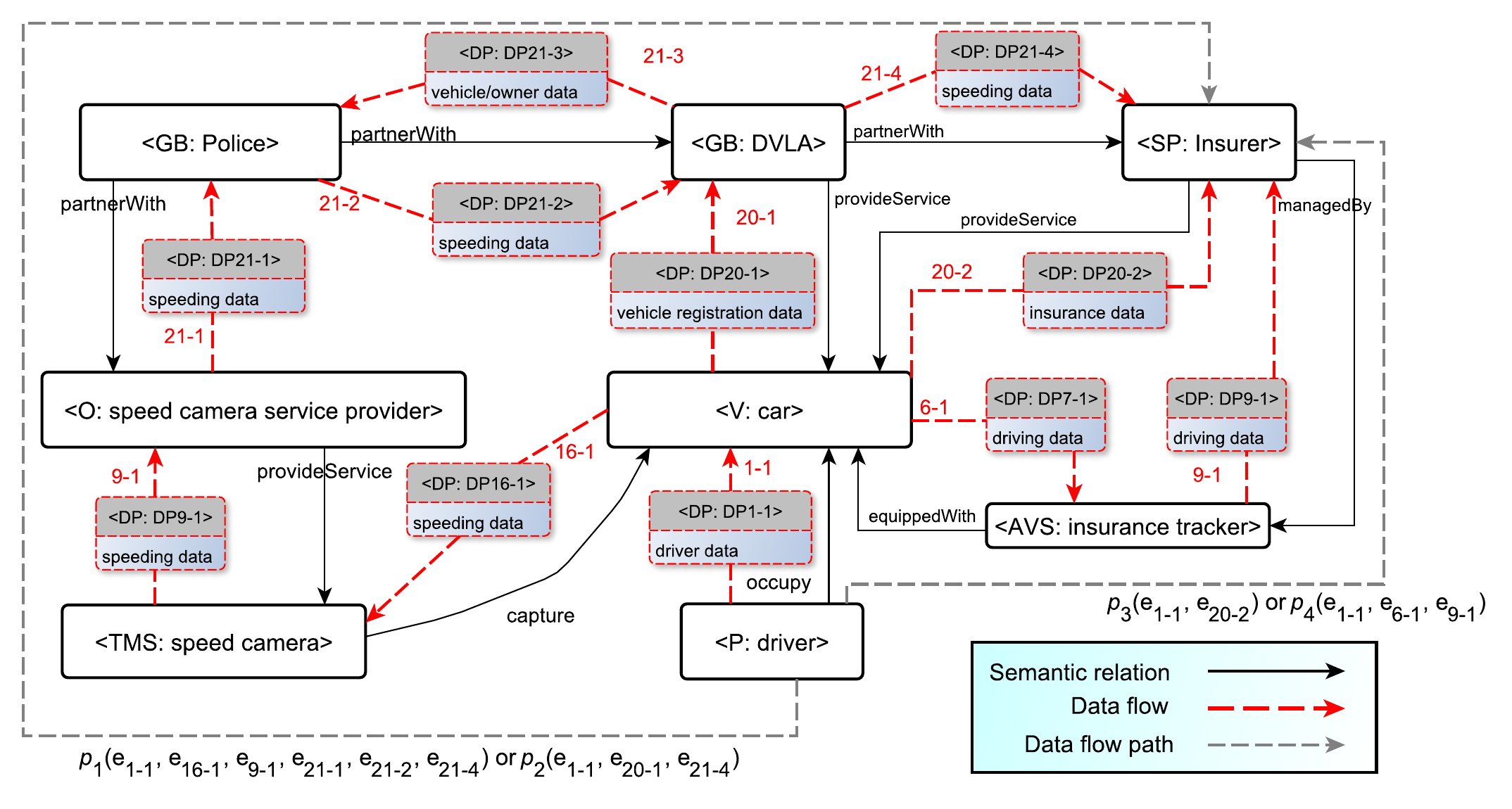}
\caption{An entity-level graph for a speeding scenario}
\label{fig:speeding_model}
\end{figure}

Here we present our work of developing an entity-level graph for modelling a speeding incident that involves GB, SP, V, TMS, and AVS entities. For clarification, we established the following assumptions: 1) the incident took place in the UK; 2) the driver owns the vehicle; 3) the driver has admitted to the speeding offence; 4) the insurance company is enrolled in the MyLicence scheme~\cite{mylicense}, enabling it to access the driver's driving history held by the UK's Driver and Vehicle Licensing Agency (DVLA); and 5) the driver has agreed to install an insurance tracker for reduced insurance premium.

\subsubsection{Modelling}

As shown in Figure~\ref{fig:speeding_model}, $e1_1$ is used to denote the flow of $<$DP1-1$>$ from the $<$driver$>$ to the $<$car$>$, where the $<$car$>$ is registered with $<$DVLA$>$ with related $<$DP20-1$>$ including the car's vechicle registration number (VRM), make and model. The data flow is modelled using edge type $E_20$, denoted by $e_{20-1}$. A $<$car$>$ is legally required to be insured, hence $<$Insurer$>$ provides insurance (i.e., \emph{provideService}) to the $<$car$>$, while $<$DP20-2$>$ insurance data has to be submitted to complete the process (i.e., denoted by $e_{20-2}$). In addition, an $<$insurance tracker$>$ installed on the car collects driving data in terms of speed, use of breaks, etc., leading to the flow of $<$DP7-1$>$ from $<$car$>$ to $<$insurance tracker$>$ (i.e., $e_{6-1}$), and subsequently $<$DP9-1$>$ is sent from $<$insurance tracker$>$ to $<$insurer$>$ (i.e., $e_{9-1}$). In a speeding scenario, $<$speed camera$>$ records $<$DP16-1$>$ such as the car's speed of travelling, the car's VRM, and images of the car and the driver. This is captured using $E_16$ edge type, denoted by $e_{16-1}$. These data records (i.e., $<$DP9-1$>$) will then be transmitted to $<$speed camera service provider$>$, denoted by $e_{9-1}$. Assume that $<$Police$>$ is \emph{partnerWith} the $<$speed cameras service provider$>$, $<$DP21-1$>$ flows to $<$Police$>$ as denoted as $e_{21-1}$ for further processing. The close partnership between $<$Police$>$ and $<$DVLA$>$ enables the flow of $<$DP21-2$>$ between the two entities, denoted as $e_{21-2}$. Based on the number plate information included in the $<$DP21-2$>$, $<$DVLA$>$ shares the relevant vehicle and its owner information (i.e., $<$DP21-3$>$) with $<$Police$>$, denoted by $e_{21-3}$. 
Furthermore, due to the participation of MyLicence scheme, $<$Insurer$>$ obtains $<$DP21-4$>$ from $<$DVLA$>$, denoted by $e_{21-4}$.

\subsubsection{Path analysis and discussion}

As illustrated in Figure~\ref{fig:speeding_model}, a driver's information could be surprisingly shared with an insurance company via four paths. The path $p_1=(e_{1-1}, e_{16-1}, e_{9-1}, e_{21-1}, e_{21-2}, e_{21-4})$ is the longest path where multiple parties are involved. Path $p_2=(e_{1-1}, e_{20-1}, e_{21-4})$ represents the case that would normally occur if an insurance company joins the MyLicence scheme. Path $p_3=(e_{1-1}, e_{20-2})$ is the shortest path that illustrates the general data sharing practice for insuring a vehicle. Path $p_4=(e_{1-1}, e_{6-1}, e_{9-1})$ describes the data sharing and collection that would happen if a driver decides to install an insurance tracker provided by an insurer. We believe that such opaque insights would not be revealed and identified without a systematic analysis using our model. Our analysis emphasises the usefulness of the entity-level graph in revealing data flow insights that might otherwise be hidden or neglected, which can lead to more potential privacy concerns and regulatory compliance needs. Moreover, when considering broader business relationships (e.g., subsidiaries and affiliated organisations) of an insurance company, the complexity of modelling such cases would increase exponentially. Although this is beyond the scope of this study, it is worth looking at possibilities to integrate with other ontologies/models that focus on business-to-business relationships for further enhancing the comprehensive understanding of the vehicle-centric data sharing ecosystem.

\subsection{More discussion on the practical usefulness}



Modern vehicles such as autonomous vehicles are equipped with a combination of sensors. The data sharing between these sensors, vehicle OBUs, and communication infrastructure is essential for driving decision-making and navigation. Our proposed model can reveal detailed insights, facilitating the analysis and identification of critical points where data latency or loss could impact the decision-making process. This, in turn, can potentially improve system reliability and safety. Additionally, in the context of smart cities, the model can illustrate and visualise the interactions between vehicles and other infrastructures, aiding urban planners in designing smarter and more responsive traffic management systems. Furthermore, considering the increasing complexity of data sharing for future transportation modes such as mobility-as-a-service, the model can potentially help pinpoint where data can be anonymised/minimised and where access control and authentication are critical. This not only ensures that user data is used only when necessary, reducing privacy risks, but also aids organisations and automotive companies in complying with regulatory requirements such as the EU's GDPR, designing better privacy policies and consent management frameworks, and implementing stronger privacy protection/preservation measures.

\section{Limitations and future work}
\label{sec:discussion}

During model development, we used various sources to identify relevant entities and relations in the ecosystem. However, there may be other useful data sources (e.g., automotive industry databases) we did not explore, which is a limitation that deserves further investigation. 
Additionally, our model does not address how data sharing changes over time. Since data sharing occurs at different times among various entities, the absence of temporal considerations could affect the accuracy of real-world scenario modelling and subsequent analyses. We acknowledge the challenges of integrating temporal information into entity models and consider this a future research direction to further develop and refine our model. Apart from the above, we have identified several other areas to enhance our proposed model and its applications. Firstly, leveraging tools such as the Web Ontology Language (OWL) and the Semantic Web Rule Language (SWRL) to formalise the proposed model can allow automated reasoning to reveal insights about privacy concerns/risks. Additionally, a more systematic analysis, employing a topological approach, could be carried out to assess the entity-level graphs' structure to discover related hidden/potential privacy concerns. Moreover, extending and integrating our model with other existing ontologies/models would enhance its comprehensiveness and applicability. Finally, the development of useful tools for visualising, comparing, and analysing related use cases would facilitate a more nuanced understanding of data sharing dynamics within modern vehicle ecosystems.

\section{Conclusions}
\label{sec:conclusions}

This paper introduces our work on developing a graph-based model for modelling the vehicle-centric data sharing ecosystem. We used different approaches, including 1) utilising GPT-4 to analyse privacy policies; 2) conducting a small-scale SLR; and 3) adopting an existing ontology, to derive key entities involved. Following the ontology development 101 methodology, we develop a graph-based model that can identify data flows for a modern vehicle in various contexts at the conceptual level. The proposed model serves as a base model for further analysis and expansion. Two realistic examples are also presented to demonstrate its flexibility and expandability in facilitating detailed examination across diverse transportation scenarios.

\bibliographystyle{IEEEtran}
\bibliography{main}

\end{document}